\documentclass{elsart5p}
\usepackage{graphics}
\usepackage{graphicx}
\usepackage{epsfig}
\usepackage{amssymb}
\usepackage{amsmath}

\newcommand{\pppi}{\mbox{$pp\to \{pp\}_{\!s\,}\pi^0$}}
\newcommand{\vpppi}{\mbox{$\pol{p}p\to \{pp\}_{\!s\,}\pi^0$}}
\newcommand{\pnpppi}{\mbox{$pn\to \{pp\}_s\pi^-$}}
\newcommand{\vpnpppi}{\mbox{$\pol{p}n\to \{pp\}_s\pi^-$}}
\newcommand{\ppppi}{\mbox{$pd\to p_{\rm sp}\{pp\}_{\!s\,}\pi^-$}}

\newcommand{\abs}{\mbox{$\pi^-{}^{3}\textrm{He}\to pnn_{\rm sp}$}}

\newcommand{\fmn}[2]{\mbox{${\textstyle \frac{#1}{#2}}$}}
\begin{document}

\begin{frontmatter}

\title{Differential cross section and analysing power of the
quasi-free $pn\to\{pp\}_s\pi^-$ reaction at 353~MeV}

\author[erlangen,dubna]{S.~Dymov},
\author[dubna]{T.~Azaryan},
\author[gatchina]{S.~Barsov},
\author[bochum,itep]{V.~Baru},
\author[ferrara]{P.~Benati},
\author[ferrara]{S.~Bertelli},
\author[ikp]{D.~Chiladze},
\author[gatchina]{A.~Dzyuba},
\author[ikp]{R.~Gebel},
\author[munster]{P.~Goslawski},
\author[ferrara]{G.~Guidoboni},
\author[ikp]{C.~Hanhart},
\author[ikp]{M.~Hartmann},
\author[ikp]{A.~Kacharava},
\author[munster]{A.~Khoukaz},
\author[dubna]{V.~Komarov},
\author[cracow]{P.~Kulessa},
\author[dubna]{A.~Kulikov},
\author[dubna]{V.~Kurbatov},
\author[ferrara]{P.~Lenisa},
\author[manchester,itep]{V.~Lensky},
\author[tbilisi]{N.~Lomidze},
\author[ikp]{B.~Lorentz},
\author[dubna,tbilisi]{G.~Macharashvili},
\author[munster]{M.~Mielke},
\author[gatchina]{S.~Mikirtytchiants},
\author[dubna,ikp]{S.~Merzliakov},
\author[ikp]{H.~Ohm},
\author[munster]{M.~Papenbrock},
\author[ikp]{F.~Rathmann},
\author[dubna,ikp]{V.~Serdyuk},
\author[dubna]{V.~Shmakova},
\author[ikp]{H.~Str\"oher},
\author[tbilisi]{M.~Tabidze},
\author[dubna]{D.~Tsirkov},
\author[rossendorf]{S.~Trusov},
\author[dubna]{Yu.~Uzikov},
\author[gatchina,ikp]{Yu.~Valdau},
\author[ucl]{C.~Wilkin\corauthref{cor1}}
\ead{cw@hep.ucl.ac.uk} \corauth[cor1]{Corresponding author.}

\address[erlangen]{Physikalisches Institut II, Universit{\"a}t
Erlangen-N{\"u}rnberg, D-91058 Erlangen, Germany }
\address[dubna]{Laboratory of Nuclear Problems, Joint Institute for Nuclear
  Research, RU-141980 Dubna, Russia}
\address[gatchina]{St. Petersburg Nuclear Physics Institute, RU-188350 Gatchina,
  Russia}
\address[bochum]{Institute of Theoretical Physics, Ruhr-Universit\"{a}t, D-44780
Bochum, Germany}
\address[itep]{Institute for Theoretical and Experimental Physics,  RU-117218 Moscow,
Russia}
\address[ferrara]{Universit\`a di Ferrara and INFN, IT-44100 Ferrara,
Italy}
\address[ikp]{Institut f\"ur Kernphysik, Forschungszentrum J\"ulich, D-52425
  J\"ulich, Germany}
\address[munster]{Institut f\"ur Kernphysik, Universit\"at M\"unster,
D-48149 M\"unster, Germany}
\address[cracow]{Institute of Nuclear Physics, PL-31342 Cracow,
Poland}
\address[manchester]{School of Physics and Astronomy, University of Manchester, Manchester M13 9PL,
UK}
\address[tbilisi]{High Energy Physics Institute, Tbilisi State University, GE-0186
Tbilisi, Georgia}
\address[rossendorf]{Institut f\"ur Kern- und Hadronenphysik,
Forschungszentrum Rossendorf, D-01314 Dresden, Germany}
\address[ucl]{Physics and Astronomy Department, UCL, London WC1E 6BT, UK}
%
%

\begin{abstract}
In order to establish links between $p$-wave pion production in
nucleon-nucleon collisions and low energy three-nucleon scattering, an
extensive programme of experiments on pion production is currently underway
at COSY-ANKE. The final proton pair is measured at very low excitation
energy, leading to an $S$-wave diproton, denoted here as $\{pp\}_{\!s}$. By
using a deuterium target we have obtained data on the differential cross
section and analysing power of the quasi-free \vpnpppi\ reaction at 353~MeV.
The spectator proton $p_{\rm sp}$ was either measured directly in silicon
tracking telescopes or reconstructed using the momentum of a detected
$\pi^-$. Both observables can be described in terms of $s$-, $p$-, and
$d$-wave pion production amplitudes. Taken together with the analogous data
on the \vpppi\ reaction, full partial wave decompositions of both processes
were carried out.
\end{abstract}

\begin{keyword}
Negative pion production; Neutron proton collisions; Amplitude
analysis

\PACS 13.75.-n   
\sep 14.40.Be    
\sep 25.40.Qa    
\end{keyword}
\end{frontmatter}
%
%

There is an extensive programme of near-threshold measurements
of $NN\to\{pp\}_{\!s\,}\pi$ at the COSY-ANKE facility of the
Forschungszentrum J\"ulich~\cite{SPIN,HAN2004}. Here the
$\{pp\}_{\!s}$ denotes a proton-proton system with very low
excitation energy, $E_{pp}$, which is overwhelmingly in the
$^{1\!}S_0$ state with antiparallel proton spins. The primary
aim of these experiments is to carry out a full amplitude
analysis which would lead to a determination of the pion
$p$-wave production strength from the $^{3\!}S_1$ initial state
that could provide links with other intermediate energy
phenomena. As part of this programme, we have already presented
data on the cross section and analysing power of the \vpppi\
reaction, which can be described in terms of $s$- and $d$-wave
pion production amplitudes~\cite{TSI2011}. We here present
analogous results for the quasi-free \vpnpppi\ reaction at an
effective beam energy around 353~MeV.

In the absence of an intense monochromatic neutron beam, the study of the
\pnpppi\ reaction is most easily carried out through quasi-free $\pi^-$
production, with a proton beam incident on a deuterium target. In order to
make a full reconstruction of a \ppppi\ event, it is necessary to measure
accurately the momentum of three of the particles in the final state.
Determinations have been made at TRIUMF of both the cross
section~\cite{DUN1998} and proton analysing power~\cite{HAH1999} by detecting
the pion together with the two fast protons. The spectator proton, $p_{\rm
sp}$, was then identified through the missing mass in the reaction and its
momentum reconstructed kinematically. Data were taken for three beam
energies, though only in the central region of pion angles. The lowest of
these energies was 353~MeV and, to allow a close comparison, we have
concentrated our attention on this energy.

In contrast to the TRIUMF data, we have extended the coverage to the whole
angular domain, which is very important in the subsequent partial wave
analysis. Furthermore, by including the possibility of detecting directly the
spectator proton or the $\pi^-$, we have obtained an independent check on
some of the systematics involved.

\begin{figure}[htb]
\centering
\includegraphics[width=\columnwidth]{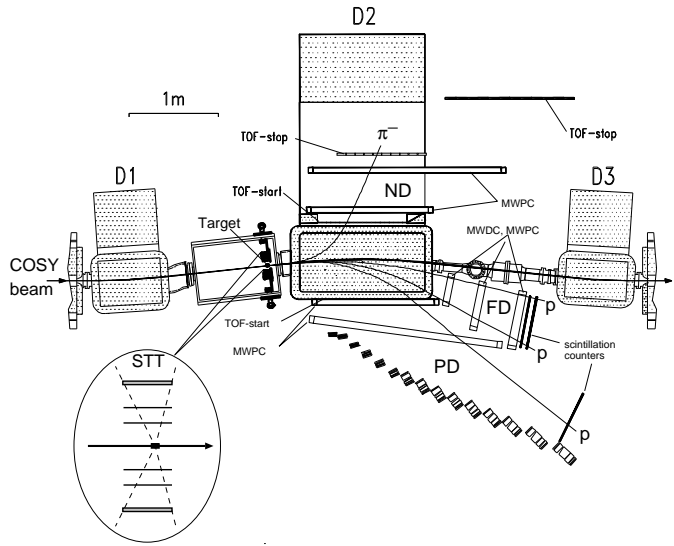}
\caption{Top view of the ANKE spectrometer setup, showing the
positions of the Positive (PD), Negative (ND), and Forward (FD)
detectors, as well as the Silicon Tracking Telescope (STT). The
dipoles D1 and D3 deflect the circulating proton beam in and out of
ANKE, whereas D2 serves as an analysing magnet.} \label{fig:AnkeFull}
\end{figure}

The experiment was carried out at the ANKE spectrometer
facility~\cite{BAR1997}, which is installed inside the COSY cooler
synchrotron storage ring of the Forschungszentrum J\"{u}lich. The circulating
proton beam was polarised perpendicularly to the horizontal plane of the
machine and the polarisation reversed in direction every six minutes.

Fast protons arising from the interaction of the beam with the deuterium
cluster-jet target~\cite{KHO1999} traversed the spectrometer dipole magnet D2
shown in Fig.~\ref{fig:AnkeFull} and entered the forward (FD) and/or positive
side (PD) detectors. Negative pions produced in the interaction at small
centre-of-mass (CM) angles, $\theta_\pi<40^\circ$, could be detected in the
negative side detector (ND), while slow spectator protons ($p_{\rm sp}$) were
observed in one of the silicon tracking telescopes (STT) that were located in
the target vacuum chamber close to the target jet~\cite{SCH2003,MUS2005}.

Each of the FD, PD, and ND detectors shown in Fig.~\ref{fig:AnkeFull}
includes both scintillation counters and multiwire proportional (MWPC) or
drift chambers (MWDC). The reconstruction of the particle trajectories from
hits in the wire chambers allowed the momenta of the ejectiles to be
evaluated. The counters were used to measure the arrival times and energy
losses required for particle identification.

The two STT were placed symmetrically to the left and right of the target
jet. Each telescope consists of three sensitive silicon layers, though only
the first two were used in this experiment. Spectator protons with the
energies $2.5<E_{\rm sp}<6$~MeV pass through the first layer and stop in the
second. The accuracy of the energy measurement, $\sigma(E_{\rm sp})/E_{\rm
sp} \simeq 10\%$, allow protons to be identified by their energy
loss~\cite{RIE1980}. Although higher energy recoil protons could be measured
by using the third STT layer, to stay within the range of applicability of
the spectator model, only events with $E_{\rm{sp}}<6$~MeV were retained. It
should, however, be noted that there is no lower limit on $E_{\rm{sp}}$ when
the $\pi^-$ is measured.

In order to provide sufficient resolution in $E_{pp}$, both fast protons from
the diproton have to be detected and, to permit a complete reconstruction of
the kinematics of the reaction, one has in addition to detect either the
$\pi^-$ or the slow spectator proton. The triggers selected fast proton
pairs, with either both protons hitting the FD or PD, or one proton recorded
in each detector. A trigger for single tracks in the FD was also added to
record deuterons from the $pn\to d\pi^0$ process, which were used for
normalisation and polarimetry purposes. A sample of data was also taken with
a special trigger requiring a coincidence between the first two layers of the
STT. This was used to investigate the STT performance and for background
studies.

The CM energy of the quasi-free $pn$ system depends on the energy and the
emission angle of the spectator and for this experiment the effective
``free'' beam energy was in the range $T_{\rm{free}}\approx
(310$--$390)$~MeV. However, for the reconstructed events, $T_{\rm{free}}$ was
measured with an accuracy of $\sigma(T_{\rm{free}})=2$--$4$~MeV and only data
in the range $T_{\rm{free}}=353\pm 20$~MeV were used in the analysis.

The identification of the \ppppi\ reaction starts from the selection of pairs
of fast protons through the difference between their times of flights in
ANKE~\cite{DYM2010}. Since the beam energy is quite low, other pairs of
particles gave a negligible contribution to the background, which originated
mostly from accidental coincidences, and amounted to $\lesssim 7\%$.  To
select the $^{1\!}S_0$ diproton state, we imposed a $E_{pp}<3$~MeV cut on the
data, which could be done reliably because of the excellent resolution of
$\sigma(E_{pp})<0.6$~MeV. It is important to note that the identical cut was
placed on the \pppi\ data~\cite{TSI2011}, which is important when considering
the relative normalisation uncertainties for $\pi^0$ and $\pi^-$ production.

The time-of-flight criterion was also used to select the $\pi^-$ in the ND,
where the very low background consisted only of accidentals and strongly
scattered positively charged particles. The background level for the
spectator protons identified in the STT was at the 5--8\% level, depending on
the proton energy and angle.

\begin{figure}[htb]
\centering
\includegraphics[width=\columnwidth]{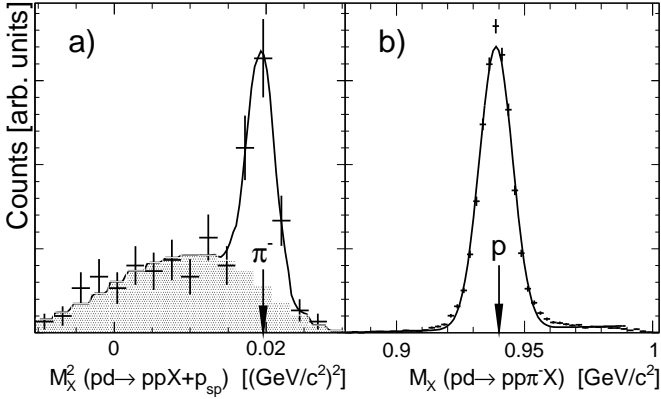}
\caption{Kinematic identification of the \ppppi\ reaction. a)
Sample of missing-mass squared data from events where the
spectator proton is detected, showing the experimental spectrum
(with error bars), the background (shaded area), and the sum of
this plus a Gaussian for the $\pi^-$ peak (solid curve). b)
Missing-mass distribution when the $\pi^-$ is detected. The
curve shows the fit to the experimental data with the sum of a
Gaussian centred on the mass of the proton and a linear
background.} \label{fig:Mx2.14}
\end{figure}

Having measured three of the final particles, the residual one
from the \ppppi\ reaction was identified by the missing-mass
method. This is illustrated separately in Fig.~\ref{fig:Mx2.14}
for the cases where the spectator proton or $\pi^-$ was
detected. The main backgrounds were from accidental
coincidences, which were particularly significant for spectator
protons because no timing information from the STT was used in
the analysis. The shape of this background was derived from
artificially constructed events. The momenta of the fast
protons were taken from the experimental events where neither a
spectator proton nor $\pi^-$ was detected, to which was added a
random spectator momentum taken from the sample acquired with
the STT trigger. The missing-mass spectrum was then fitted with
the sum of this background distribution and a Gaussian. The
background level was estimated separately for each detector
combination, spin orientation of the beam, and angular bin. Any
uncertainties here were combined with the statistical errors.

In order to estimate the resolutions and acceptance of the system, a full
simulation of the ANKE setup was performed~\cite{DYM2010}, based on the GEANT
package~\cite{AGO2003}. The acceptance was calculated as a function of
$\theta_\pi$ for both the $\pi^-$ and spectator proton detection, assuming
that the spectator was emitted isotropically, with the energy being sampled
from the Fermi distribution~\cite{FRO2007}. The fast proton pair was
generated in the $^{1\!}S_0$ state, with the distribution in excitation
energy being weighted by the Migdal-Watson factor~\cite{WAT1952} that
included the Coulomb interaction~\cite{DYM2006}.

The polarisation of the proton beam and the luminosity were both estimated
from quasi-free $\pol{p}n\to d\pi^0$ data that were taken in parallel. The
fast deuteron was detected in the FD, and selected by its energy loss in the
counters, and the spectator proton measured in coincidence in the STT. The
reaction was then identified by the missing-mass method.

The measurements of both the analysing power and cross section depend
sensitively on the relative luminosities for the two spin states and this was
controlled by comparing the rates of ejectiles emitted at $\theta=0^\circ$ or
$\phi=\pm90^\circ$, which cannot depend upon the beam polarisation. Using
calibration data taken from the SAID database~\cite{SAID}, the integrated
luminosity was found to be $L=(2310\pm110)~\textrm{nb}^{-1}$. No correction
was made for the shadowing in the deuteron since there is a similar effect in
the measurement of the quasi-free \pnpppi\ rate. The beam polarisation
determined in this way was $P=0.66\pm0.06$. This is consistent with that
found in the $\pi^0$ production experiment~\cite{TSI2011} that was undertaken
with the same conditions just before the $\pi^-$ run. The luminosities in the
two polarisation states were on average very close, $L_\uparrow/L_\downarrow
= 1.017 \pm 0.005$.

\begin{figure}[htb]
\vspace*{1mm}
\centering
\includegraphics*[width=\columnwidth]{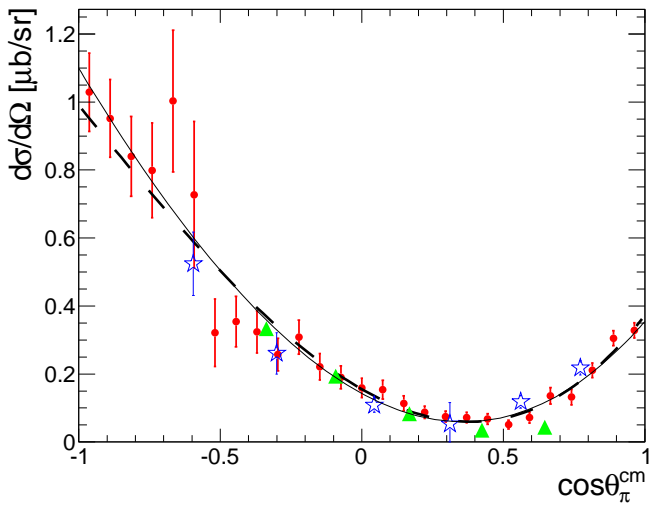}

\caption{Unpolarised differential cross section for the \pnpppi\ reaction at
$T_{\rm free}\approx 353$~MeV. The ANKE data with statistical errors are
shown by red circles. In addition there is a systematic error of 6\% arising
from the luminosity and acceptance determination. The statistical errors of
the TRIUMF \pnpppi\ results~\cite{DUN1998} (green triangles) are generally
smaller than the symbol size and their normalisation uncertainty is 10\%. The
arbitrarily scaled TRIUMF cross sections extracted from \abs\
data~\cite{HAH1996} (blue stars) are also included. The dashed curve is a
direct cubic fit to these ANKE data whereas the solid one corresponds to the
global fit described in the text.} \label{fig:DSG}
\end{figure}

Figure~\ref{fig:DSG} shows the differential cross-section $d\sigma/d\Omega$
for the quasi-free \pnpppi\ reaction integrated over the $E_{pp}=(0-3)$~MeV
range of $pp$ excitation energies and averaged over the effective beam energy
$T_{\rm{free}}=353\pm 20$~MeV. The data were extracted within the impulse
approximation model, where the weighting of the spectator momentum
distribution was taken from the Bonn deuteron wave function~\cite{MAC1987},
though the result was insensitive to this choice. The cross-section was
estimated separately for $\pi^-$ and spectator proton detection and, since
these two results were found to be consistent, only their weighted average is
presented. Also shown are the TRIUMF data on quasi-free $\pi^-$
production~\cite{DUN1998}. They imposed a slightly more severe $E_{pp}$ cut
on their data and these were converted to our 3~MeV cut by assuming a
Migdal-Watson energy variation~\cite{WAT1952}.

Whereas the TRIUMF results only cover the central region of pion
angles~\cite{DUN1998}, the current data extend over the whole angular domain.
The two data sets are consistent in the backward hemisphere but the TRIUMF
measurements show no indication of the rise at forward angles that is seen at
ANKE. Some confirmation of the ANKE angular shape is offered by pion
absorption data, \abs, where the unobserved slow neutron is assumed to be a
spectator~\cite{HAH1996}. In this case the reaction can be interpreted as
being $\pi^-\{pp\}_{\!s}\to pn$, though the internal structure of the
diproton is very different to that in the production data. Over the range of
angles covered, our data are completely consistent with these absorption
results. The forward/backward peaking is in complete contrast to the results
found for $\pi^0$ production~\cite{BIL2001,TSI2011} and is an indication of
the dominance of the $I=0$ $p$-wave amplitudes in this reaction.

The unpolarised cross section for $\pi^-$ production, and this times the
proton analysing power $A_y$, must be of the form
\begin{eqnarray}
\label{dsig}
\left(\frac{d\sigma}{d\Omega}\right)_{\!0}&=&\frac{k}{4p}\sum_{n=0}
a_n\cos^n\theta_{\pi},\\
\label{aydsig}
A_y\left(\frac{d\sigma}{d\Omega}\right)_{\!0}&=&\frac{k}{4p}
\sin\theta_{\pi}\sum_{n=0}b_{n+1}\cos^n\theta_{\pi},
\end{eqnarray}
where $\theta_{\pi}$ is the pion c.m.\ production angle with respect to the
direction of the polarised proton beam. Here $p$ is the incident c.m.\
momentum and $k$ that of the produced pion. We are neglecting here any small
effects due to the mass differences and at 353~MeV; the momenta then have
values $p=407$~MeV/$c$ and $k\approx 94$~MeV/$c$. The best fits of
Eq.~\eqref{dsig} to the differential cross section are found with the
parameters quoted in Table~\ref{obs_val}.

\begin{table}[htb]
\begin{center}
\begin{tabular}{|c|c|c|}
\hline
Observable&Direct fit&Global fit\\
\hline
$a_0(pp)$ &  $\phantom{-}4.05\pm0.08$ &   $\phantom{-}4.05\pm0.08$\\
$a_2(pp)$ & $-2.31\pm0.14$ & $-2.34\pm0.14$ \\
$b_2(pp)$ &  $\phantom{-}1.82\pm0.10$ &   $\phantom{-}1.80\pm0.10$ \\
$a_0(pn)$ &  $\phantom{-}2.69\pm0.18$ &  $\phantom{-}2.47\pm0.08$ \\
$a_1(pn)$ & $-8.24\pm0.51$ & $-7.83\pm0.45$ \\
$a_2(pn)$ &  $\phantom{-}9.11\pm0.70$ &  $\phantom{1}10.12\pm0.41$ \\
$a_3(pn)$ &  $\phantom{-}2.89\pm0.90$ &  $\phantom{-}1.38\pm0.27$ \\
$b_1(pn)$ &  $\phantom{-}1.77\pm0.14$ &  $\phantom{-}1.82\pm0.13$  \\
$b_2(pn)$ & $-1.95\pm0.50$ & $-1.75\pm0.36$ \\
$b_3(pn)$ & $-4.43\pm0.70$ & $-4.83\pm0.27$ \\
\hline
\end{tabular}
\caption{Values of the parameters in $\mu$b/sr extracted by
direct fits of Eqs.~\eqref{dsig} and \eqref{aydsig} to the
\pppi\ and \pnpppi\ experimental data and those obtained on the
basis of Eq.~\eqref{relation} from the amplitudes given in
Eq.~\eqref{amp_vals}. The error bars are purely statistical. In
the $\pi^-$ case there are systematic uncertainties of 6\% in
the cross section and 9\% in the analysing powers. }
\label{obs_val}
\end{center}
\end{table}

\begin{figure}[!ht]
\centering
\includegraphics[width=0.95\columnwidth]{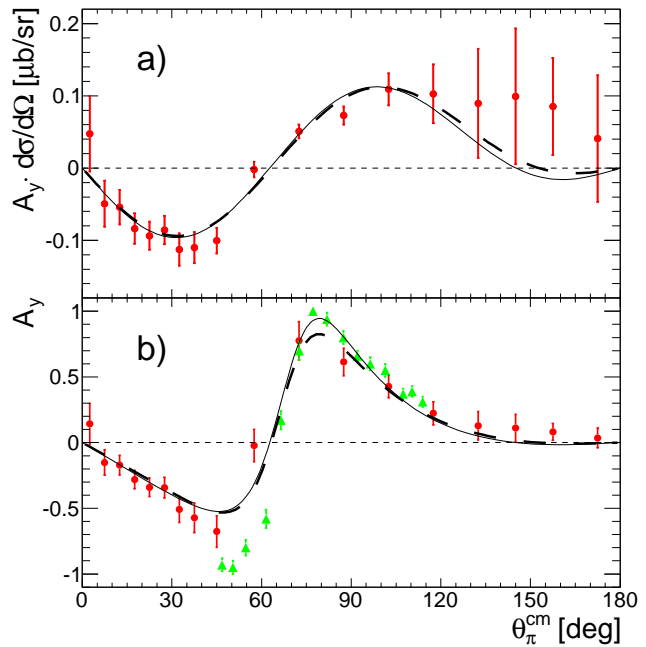}
\caption{\label{fig:DSGAy} (a) The product of the measured analysing power
and differential cross section for the \vpnpppi\ reaction at 353~MeV, where
the error bars shown are statistical and do not include the 11\% systematic
uncertainty. The dashed curve represents the best fit of Eq.~(\ref{aydsig})
with $b_1$, $b_2$, and $b_3$ terms whereas the solid one corresponds to the
global fit to all the data. (b) Measured values of $A_y$ for the \vpnpppi\
reaction showing both the ANKE (circles) and TRIUMF data~\cite{HAH1999}
(triangles). The systematic uncertainty in the ANKE data is 9\%. The lines
represent the quotients of the fits in panel-a and those to the cross section
in Fig~\ref{fig:DSG}.}
\end{figure}

The results for the analysing power of the \vpnpppi\ reaction are displayed
in Fig.~\ref{fig:DSGAy}, with $A_y(d\sigma/d\Omega)$ being shown in panel a
and $A_y$ in panel b. The agreement with the TRIUMF $A_y$ data~\cite{HAH1999}
is reasonable at large angles and both show the strong and rather asymmetric
oscillation in the central region. However, there are clear discrepancies for
$\theta_{\pi}\lesssim 60^{\circ}$, as there are also for the cross section
shown in Fig.~\ref{fig:DSG}.

Fitting the weighted $A_y$ distribution with the form of Eq.~\eqref{aydsig}
requires at least the three terms given in Table~\ref{obs_val}. The
associated curve is shown in Fig.~\ref{fig:DSGAy}a and this divided by the
parameterisation of the cross section in panel b.

The spin structure of the \pnpppi\ reaction is identical to that of \pppi,
which was already presented in~\cite{TSI2011}. The cross section and
analysing power can be written in terms of two scalar amplitudes $A$ and $B$
through
\begin{eqnarray}
\nonumber\left(\frac{d\sigma}{d\Omega}\right)_{\!0}&=&\frac{k}{4p}\left(|A|^2+|B|^2
+2\,\textrm{Re}[AB^*]\cos\theta_{\pi}\right)\!,\\
\label{obs1}
A_y\left(\frac{d\sigma}{d\Omega}\right)_{\!0}&=&\frac{k}{4p}\left(
2\,\textrm{Im}[AB^*]\sin\theta_{\pi}\right)\!.
\end{eqnarray}

Keeping terms up to pion $d$ waves, the \pppi\ data at 353~MeV~\cite{TSI2011}
can be parameterised in terms of the three amplitudes $M_s^P$, $M_d^P$, and
$M_d^F$, corresponding to the transitions, $^{3\!}P_0\to\, ^{1\!}S_0s$,
$^{3\!}P_2\to\, ^{1\!}S_0d$, and $^{3\!}F_2\to\, ^{1\!}S_0d$, respectively.
In proton-neutron collisions there are also the two $p$-wave transitions,
$^{3\!}S_1\to\, ^{1\!}S_0p$ and $^{3\!}D_1\to\, ^{1\!}S_0p$ that arise in the
isospin $I=0$ case, and for these we introduce amplitudes $M_p^S$ and
$M_p^D$, respectively.

In terms of these partial waves, $A$ and $B$ become
\begin{eqnarray}
\nonumber
A&=&\frac{1}{\sqrt{2}}\left[M_s^P-\fmn{1}{3}M_d^P+M_d^F\left(\cos^2\theta_{\pi}-\fmn{1}{5}\right)+M_p^D\cos\theta_{\pi}\right],\\
\label{amps2} B&=&\frac{1}{\sqrt{2}}\left[M_p^S-\fmn{1}{3}M_p^D+\left(M_d^P-\fmn{2}{5}M_d^F\right)\cos\theta_{\pi}\right].
\end{eqnarray}

It is clear from the very different behaviour of the angular distributions
for $\pi^0$ and $\pi^-$ production that the extra $p$-wave amplitudes in
Eq.~\eqref{amps2} must be very large and that it will not be justified to
discard their interference with $M_d^P$.

Keeping terms up to $p$--$d$ interference but omitting the squares of the
$d$-wave amplitudes, we find that
\begin{eqnarray}
\nonumber
a_0&=&\fmn{1}{2}|{M}^{P}_{\rm s}|^2+\fmn{1}{2}|{M}^{S}_{\rm p}- \fmn{1}{3}{M}^{D}_{\rm p}|^2-
\fmn{1}{3} \mathrm{Re}\left[{M}^{P*}_{\rm s}({M}^{P}_{\rm d} +\fmn{3}{5}{M}^{F}_{\rm d})\right]\\\nonumber
a_1&=&\mathrm{Re}\left[{M}^{P*}_{\rm s}({M}^{S}_{\rm p} +\fmn{1}{3}{M}^{D}_{\rm p}) +
\fmn{2}{3} {M}^{P*}_{\rm d}({M}^{S}_{\rm p} -\fmn{5}{6}{M}^{D}_{\rm p})\right.\\\nonumber
&&\left.-\fmn{3}{5} {M}^{F*}_{\rm d}{M}^{S}_{\rm p} \right]\\\nonumber
a_2&=&\fmn{1}{6}|{M}^{D}_{\rm p}|^2 +
\mathrm{Re}\left[{M}^{P*}_{\rm s}({M}^{P}_{\rm d} +\fmn{3}{5}{M}^{F}_{\rm d}) +
{M}^{S*}_{\rm p}{M}^{D}_{\rm p} \right]\\\nonumber
a_3&=& \mathrm{Re}\left[[{M}^{D*}_{\rm p}({M}^{P}_{\rm d} +\fmn{4}{15}{M}^{F}_{\rm d}) +
{M}^{S*}_{\rm p}{M}^{F}_{\rm d} \right]\\\nonumber
b_1&=& \mathrm{Im}\left[({M}^{S*}_{\rm p}- \fmn{1}{3}{M}^{D*}_{\rm p}) ({M}^{P}_{\rm s}
-\fmn{1}{3}({M}^{P}_{\rm d} +\fmn{3}{5}{M}^{F}_{\rm d}))\right]\\\nonumber
b_2&=& \mathrm{Im}\left[{M}^{S*}_{\rm p}{M}^{D}_{\rm p}-
{M}^{P*}_{\rm s}({M}^{P}_{\rm d} -\fmn{2}{5}{M}^{F}_{\rm d}) \right]\\
b_3&=& \mathrm{Im}\left[{M}^{P*}_{\rm d}{M}^{D}_{\rm p}-
{M}^{F*}_{\rm d} ({M}^{S}_{\rm p} +\fmn{1}{15}{M}^{D}_{\rm p}) \right] \ .
\label{relation}
\end{eqnarray}

By neglecting the small coupling between the $^{3\!}P_2$ and $^{3\!}F_2$
partial waves, and imposing the Watson theorem to determine the phases, it
was possible to extract values for the complex amplitudes $M_s^P$, $M_d^P$,
and $M_d^F$ from the analysis of the \pppi\ reaction~\cite{TSI2011}. Such an
approach would not be valid for the two $p$-wave terms because of the very
strong coupling between the incident $^{3\!}S_1$ and $^{3\!}D_1$ waves.
Nevertheless, although there is a significant overall relative uncertainty
between the $\pi^-$ and $\pi^0$ production data, associated with luminosity
and other systematic effects, it is clear from the comparison of the seven
free parameters in Eq.~\eqref{relation} with the ten observables in
Table~\ref{obs_val} that the system is over determined.  As a consequence, if
an acceptable solution is achieved it would support the approximation made in
our analysis, such as the neglect of higher partial waves, $d$-$d$
interference and the effect of coupling between the $^{3\!}P_2$ and
$^{3\!}F_2$ partial waves

The best fit to the combined \pppi\ and \pnpppi\ data sets is obtained with
\begin{eqnarray}
\nonumber
M_s^P &=& \phantom{-}(55.3\pm0.4)-(14.7\pm0.1)i~\sqrt{\textrm{nb/sr}},\\\nonumber
M_d^P &=& -(26.6\pm1.1)-(8.6\pm0.4)i~\sqrt{\textrm{nb/sr}},\\\nonumber
M_d^F &=& \phantom{-(}5.3\pm 2.3~\sqrt{\textrm{nb/sr}},\\\nonumber
M_p^S &=& -(32.4\pm2.2)+(17.3\pm2.7)i~\sqrt{\textrm{nb/sr}},\\
M_p^D &=& -(109.6\pm9.6)+(140.7\pm4.0)i~\sqrt{\textrm{nb/sr}}.
\label{amp_vals}
\end{eqnarray}
Since this solution has $\chi^2/\textrm{NDF}=89/82$, it shows that our
truncated expansions can give a very good description of the data. The
contribution from the $^{3\!}F_2\to\, ^{1\!}S_0d$ transition is clearly very
small and, if one eliminated $M_d^F$ completely, it would give only a
marginally poorer fit with $\chi^2/\textrm{NDF}=94/82$. Note, whereas the
phases of $M_s^P$, $M_d^P$ and $M_d^F$ were imposed (see above), the phases
of $M_p^S$ and $M_p^D$ were extracted from the data.

The quality of the fits can also be judged from the comparison of the curves
in Figs.~\ref{fig:DSG} and \ref{fig:DSGAy} with the data. The residual small
discrepancies in the description of the analysing power might, of course, be
due to the neglect of some of the smaller terms. However, it should also be
borne in mind that the main systematic uncertainty, namely the relative
normalisations between the \pppi\ and \pnpppi\ data sets, has not been
included in the determination of the parameters. On the other hand, adjusting
this by a few percent would not lead to any qualitative changes in the
solution of Eq.~\eqref{amp_vals}.

Another way of judging the changes introduced by making a global fit to all
the data simultaneously rather than fits to individual distributions is to
look by how much the parameters themselves were changed by this procedure.
These are shown in Table~\ref{obs_val}. No changes at all are to be noticed
for the \pppi\ reaction and for the \pnpppi\ case it is only $a_3(pn)$ where
the difference is greater than the error bars.

The conclusions that one can draw from Eq.~\eqref{amp_vals} are first that,
although $d$-wave pion production is significant, this is almost exclusively
from the $^{3\!}P_2$ state since the $^{3\!}F_2\to\, ^{1\!}S_0d$ transition
is very weak. In the \pnpppi\ case the amplitudes are dominated by the
$^{3\!}D_1\to\, ^{1\!}S_0p$ transition.

A partial wave decomposition of the TRIUMF \vpnpppi\ data was
attempted~\cite{HAH1999} but at the time there were no $\pi^0$ production
results available to constrain the fit and the authors had to rely on the
application of the Watson theorem even for the $^{3\!}S_1$ and $^{3\!}D_1$
waves, which are strongly coupled. Furthermore, their data did not extend
over the whole angular domain and so it is not surprising that their partial
wave results bear little relation to ours. They found negligible $d$-wave
production and did not identify the dominant role played by the
$^{3\!}D_1\to\, ^{1\!}S_0p$ transition.

In summary, we have measured the differential cross section and proton
analysing power of $\pi^-$ production in the quasi-free \vpnpppi\ reaction in
the 353~MeV energy region over the whole angular domain. These complement the
analogous data obtained on $\pi^0$ production in the \vpppi\ reaction.
Through a careful use of the Watson theorem, an amplitude analysis of the
combined data sets has been achieved and these have allowed the determination
of the partial waves up to $\ell_{\pi}=2$.

Some of the biggest uncertainties in the global amplitude analysis arise from
the normalisations in the \pppi\ and \pnpppi\ data, which affect primarily
the $s$- and $p$-wave production, respectively. The relative normalisation
can be determined independently through a measurement of the transverse spin
correlation parameter $A_{x,x}$ for the \pnpppi\ reaction~\cite{DYM2010a}.
Such data have already been taken and, when the results are available in
2012, these will make the amplitude analysis more robust.

The pion-production amplitudes extracted must now be analysed using chiral
perturbation theory and this will be an important step to provide a deeper
understanding of low energy pion dynamics. If this programme is successful,
it will provide further strong evidence that chiral perturbation theory can
indeed be applied to $NN\to NN\pi$ provided that the large momentum transfers
are taken into account properly. This is a precondition for the analysis of
isospin-violating reactions in the same
framework~\cite{FIL2009,GAR2004,NOG2006}, which will shed light on the role
of the quark mass term in low energy pion reactions.

We are grateful to other members of the ANKE Collaboration for
their help with this experiment and to the COSY crew for
providing such good working conditions, especially of the
polarised beam. This work has been partially supported by the
BMBF (grant ANKE COSY-JINR), RFBR (09-02-91332), DFG (436 RUS
113/965/0-1), the JCHP FFE, the SRNSF (09-1024-4-200), the
Helmholtz Association (VH-VI-231), STFC (ST/F012047/1 and
ST/J000159/1), and the EU Hadron Physics 2 project ``Study of
strongly interacting matter''
%
%

\end{document}